# The effects of different quantum feedback types on the tightness of the variance-based uncertainty

Xiao Zheng(郑晓)[1], Guo-Feng Zhang(张国锋)[1,2,*]

[1]*Key Laboratory of Micro-Nano Measurement-Manipulation and Physics (Ministry of Education), School of Physics and Nuclear Energy Engineering, Beihang University, Xueyuan Road No. 37, Beijing 100191, China*

[2]*State Key Laboratory of Low-Dimensional Quantum Physics, Tsinghua University, Beijing 100084, China*

**Abstract:** The effect of the quantum feedback on the tightness of the variance-based uncertainty, the possibility of using quantum feedback to prepare the state with a better tightness, and the relationship between the tightness of the uncertainty and the mixedness of the system are studied. It is found that the tightness of Schrodinger-Robertson uncertainty (SUR) relation has a strict liner relationship with the mixedness of the system. As for the Robertson uncertainty relation (RUR), we find that the tightness can be enhanced by tuning the feedback at the beginning of the evolution. In addition, we deduce that the tightness of RUR has an inverse relationship with the mixedness and the relationship turns into a strict linear one when the system reach the steady state.

**PACS number(s)**: 03.65.Yz; 03.65.Ud; 42.50.Pq

**Keywords:** quantum feedback; uncertainty; mixedness; tightness

## I. INTRODUCTION

The uncertainty principle which is similar to the quantum entanglement is one of the most remarkable characters of quantum mechanics as well as a fundamental departure from the principle of classical physics [1-7]. Any pair of incompatible observables complies with a certain form of uncertainty relationship, the constraint of which sets the ultimate bound on the measurement precision achievable for these quantities. The conventional variance-based uncertainty relations which is first deduced by Heisenberg in the circumstance of position and momentum possess a clear physical conception and still find a variety of applications in quantum information science, such as entanglement detection [4,8], quantum spin squeezing [9–14], and even quantum metrology [15–17]. Therefore the improvement of the tightness and the lower bound of the uncertainty becomes very important [18]. Robertson uncertainty relation (RUR) is the most famous form among them, which reads [2]:

$$(\Delta A)^2(\Delta B)^2 \geq \left|\frac{1}{2i}\langle[A,B]\rangle\right|^2, \tag{1}$$

where the standard deviation $\Delta O$ and expectation value $\langle O \rangle$ are taken over the state $\rho$ with $O \in \{A, B\}$. Meanwhile, it is worth noting that the RUR can be derived from another strengthened inequality, the Schrodinger-Robertson uncertainty relation (SUR) [19, 20]:

$$(\Delta A)^2(\Delta B)^2 \geq \left|\frac{1}{2i}\langle[A,B]\rangle\right|^2 + \left|\frac{1}{2}\langle\{\check{A},\check{B}\}\rangle\right|^2. \tag{2}$$

[*] Correspondence and requests for materials should be addressed to G.Z.(gf1978zhang@buaa.edu.cn)

Here $I$ is the identical operator and $\check{O} = O - \langle O \rangle I$. The uncertainty inequalities of the types of Eq. (1) and Eq. (2) are often known as Heisenberg-type and Schrodinger-type uncertainty relations, respectively.

Quantum feedback [21], as a means of controlling the system, is more and more valued by researchers in the preparation of the special state. Thus it would be of great interest to investigate the effect of the quantum feedback control on the tightness of the uncertainty. In this paper we mainly study the influence of the feedback on the properties of RUR and SUR, and the relationship between the tightness of the uncertainty and the mixedness of the system. An outline of the paper is as follows. In Sec. II, the physical model a qubit interacting with a dissipative cavity which provide the feedback to the qubit is introduced. The effects of the feedback on the tightness and mixedness are studied in Sec. III, Finally, Sec. IV is devoted to the discussion and conclusion.

## II. THE PHYSICAL MODEL

The physical model of a single atom resonantly coupled to a single-mode cavity, which is damped with decay rate $\kappa$, will be introduced. As shown in Fig. 1.

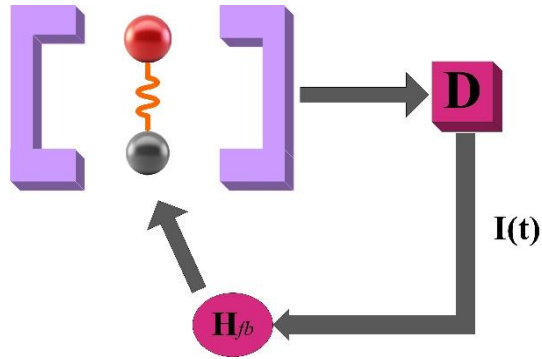

Fig. 1: Schematic view of the model, the feedback Hamiltonian is applied to the atoms according to the homodyne current I(t) derived from detector D.

In the model, we will consider the Markovian feedback [21], with the control Hamiltonian $H_{fb} = I(t)F$, where $I(t)$ is the signal from the homodyne detection of the cavity output. In the homodyne-based scheme, the detector registers a continuous photocurrent and the feedback Hamiltonian is constantly applied to the system. The master equation is read as [22, 23]

$$\frac{d\rho}{dt} = -i\left[\frac{1}{2}(\sigma^+ F + F\sigma_-), \rho\right] + D(\sigma_- - iF)\rho, \qquad (3)$$

where $\rho$ is the density matrix of the qubit, and $D(\mathcal{A})\rho = \mathcal{A}\rho\mathcal{A}^+ - (\mathcal{A}^+\mathcal{A}\rho + \rho\mathcal{A}^+\mathcal{A})/2$ represents the irreversible evolution induced by the interaction between the system and the environment.

### III. THE EFFECT OF FEEDBACK ON UNCERTAINTY

An arbitrary Hermitian operator of qubit systems can be denoted by $\varepsilon_1 I + \varepsilon_2 \vec{a} \vec{\sigma}$, where $\{\varepsilon_1, \varepsilon_2\}$ are real parameters, $\vec{a} \in R_3$ are unit vectors, and $\vec{\sigma} = (\sigma_x\ \sigma_y\ \sigma_z)$ are standard Pauli matrices. If we choose the identity operator $I$ as feedback, the system will not be affected by tuning the feedback due to the properties of identity operator. Therefore, in order not to lose generality, two different Hermitian operators $\lambda(\sin(\beta)\sigma_x + \cos(\beta)\sigma_y)$ and $\mu\sigma_z$ ($\lambda \in (-1,1), \mu \in (-1,1)$, and $\beta \in (0,2\pi)$) are selected as the feedback operators to study the effect of different feedback types on the uncertainty. In order to research the tightness of the RUR and SUR, we define:

$$U = \Delta\hat{A}^2\Delta\hat{B}^2 - \frac{1}{4}\left|\langle[\hat{A},\hat{B}]\rangle\right|^2 \qquad (4)$$

$$W = \Delta\hat{A}^2\Delta\hat{B}^2 - \frac{1}{4}\left|\langle[\hat{A},\hat{B}]\rangle\right|^2 - \left|\frac{1}{2}\langle\{\check{A},\check{B}\}\rangle\right|^2 \qquad (5)$$

It is easy to see that the smaller the value of $U(W)$ is, the better the tightness of RUR (SUR) is. For the density matrix $\rho$, the state is a pure one when $\text{Tr}(\rho^2) = 1$, and $\text{Tr}(\rho^2) < 1$ for the mixed one. Denoting $1 - \text{Tr}(\rho^2)$ by $\Upsilon$, therefore the value of $\Upsilon$ can be employed to detect the mixedness of qubits states. We can deduce that the bigger the value of $\Upsilon$ is the bigger the mixedness of $\rho$ is. In the following, we mainly focus on the effect of feedback and mixedness on the uncertainty relation from the aspect of tightness. We take $\hat{A} = \sigma_x$ $\hat{B} = \sigma_z$ and choose superposition state $|\varphi_0\rangle = \cos(\alpha)|g\rangle + \sin(\alpha)|e\rangle$ as the initial state in the following.

#### A. The feedback $\lambda(\sin(\beta)\sigma_x + \cos(\beta)\sigma_y)$

The feedback $\lambda(\sin(\beta)\sigma_x + \cos(\beta)\sigma_y)$ will be studied in this subsection. Substituting $F = \lambda(\sin(\beta)\sigma_x + \cos(\beta)\sigma_y)$ into Eq. (3), one can obtain the density matrix:

$$\rho(t) = \begin{pmatrix} \rho_{11}(t) & \rho_{12}(t) \\ \rho_{12}{}^*(t) & 1 - \rho_{11}(t) \end{pmatrix} \tag{6}$$

with the elements

$$\rho_{11}(t) = \frac{(1 + 2e^{\mathcal{T}t}\lambda^2 - (1 + 2\lambda^2)\cos(2\alpha) + 4\lambda\cos(\beta)\sin(\alpha)^2)}{2\mathcal{T}e^{\mathcal{T}t}} \tag{7}$$

$$\rho_{12}(t) = -\frac{\sin(2\alpha)(-2e^{i\beta}\xi + 2ie^{2t\lambda\mathcal{P}}(1 + e^{i\beta}\lambda)\sin(\beta))}{4\sqrt{\mathcal{O}}\mathcal{P}} \tag{8}$$

in which $\mathcal{O} = \exp(t + 4t\lambda^2 + 4t\lambda\cos(\beta))$, $\mathcal{P} = \lambda + \cos(\beta)$, $\mathcal{T} = 1 + 2\lambda^2 + 2\lambda\cos(\beta)$, $\xi = 1 + \lambda\cos(\beta)$ After a simple calculation, we have

$$\left(\Delta \hat{A}\right)^2 = 1 - \frac{\sin(2\alpha)^2(2\cos(\beta) + \lambda(1 + \cos(2\beta) + 2e^{2t\lambda\mathcal{P}}\sin(\beta)^2))^2}{4\mathcal{O}\mathcal{P}^2} \tag{9}$$

$$\left(\Delta \hat{B}\right)^2 = 1 - \frac{[\mathcal{T}\cos(2\alpha) + (e^{\mathcal{T}t} - 1)(1 + 2\lambda\cos(\beta))]^2}{\mathcal{T}^2 e^{2\mathcal{T}t}} \tag{10}$$

$$\left|\langle[\hat{A}, \hat{B}]\rangle\right|^2 = \frac{4(1 + \lambda\cos(\beta))^2 \sin(2\alpha)^2 \sin(\beta)^2 (e^{2\mathcal{P}t\lambda} - 1)^2}{\mathcal{O}\mathcal{P}^2} \tag{11}$$

$$\left|\langle\{\breve{A}, \breve{B}\}\rangle\right|^2 = \frac{4((e^{\mathcal{T}t} - 1)(1 + 2\lambda\cos(\beta)) + \cos(2\alpha)\mathcal{T})^2 \sin(2\alpha)^2 \mathcal{G}^2}{\mathcal{P}^2 \mathcal{T}^2 e^{3\mathcal{T}t}} \tag{12}$$

$$\Upsilon = \frac{(-\mathcal{P}^2 \ell(\ell - 2\mathcal{T}e^{\mathcal{T}t}) - \mathcal{T}^2 e^t \sin(2\alpha)^2 (\xi^2 + e^{2\mathcal{P}t\lambda}(-2\xi + e^{2\mathcal{P}t\lambda}(\mathcal{T} - \lambda^2))\sin(\beta)^2))}{2e^{2\mathcal{T}t}\mathcal{T}^2 \mathcal{P}^2} \tag{13}$$

with $\ell = 1 + 2\lambda^2 \exp(\mathcal{T}t) - (1 + 2\lambda^2)\cos(2\alpha) + 4\lambda\cos(\beta)\sin(\alpha)^2$ and $\mathcal{G} = \mathcal{P}\cosh(t\lambda\mathcal{P}) - (\cos(\beta) + \lambda\cos(2\beta))\sinh(t\lambda\mathcal{P})$. According to the above formulas, one can deduce that:

$$W = \Delta\hat{A}^2 \Delta\hat{B}^2 - \frac{1}{4}\left|\langle[\hat{A}, \hat{B}]\rangle\right|^2 - \left|\frac{1}{2}\langle\{\breve{A}, \breve{B}\}\rangle\right|^2 = 2\Upsilon \tag{14}$$

which means the tightness of the SUR has a strict linear relationship with the mixedness of the system and the less the mixedness is the better the tightness of the SUR is. Therefore we can investigate the tightness of the SUR by the evolution of the mixedness.

In the following, the influence of the feedback on the tightness of the RUR and the mixedness of the system will be investigated and we firstly concentrate on the feedback $\lambda\sigma_x$, namely take $\beta = \pi/2$. Making use of Eqs. (9), (10), (11) and (13), one can obtain the evolution of $U$ and $\Upsilon$ with respect to time for different initial state in Fig.2, here we take $\lambda = 1$.

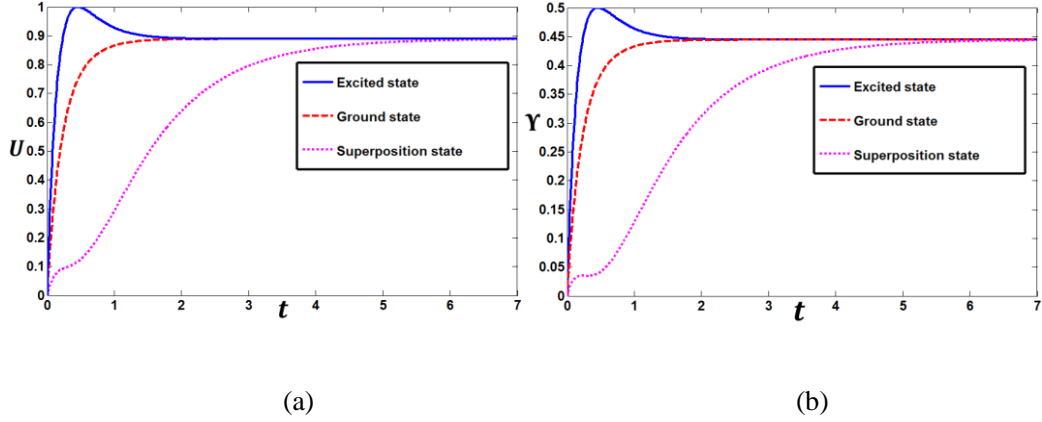

(a)                                          (b)

Fig.2. the evolution of $U$ in (a) and $\Upsilon$ in (b) with respect to $t$ for different initial state, here excited state stands for $|e\rangle$, ground state stands for $|g\rangle$ and superposition state stands for $(|g\rangle + |e\rangle)/\sqrt{2}$, here $\beta = \pi/2$.

It is easy to see from Fig.2 that choosing the superposition state $(|g\rangle + |e\rangle)/\sqrt{2}$ can be more effective to enhance the tightness of uncertainty and reduce the mixedness than choosing excited and ground one as initial state. In addition, it's easy to find that evolution of mixedness has almost the same structure with the one of $U$, which means there is a proportional relationship between the value of $U$ and $\Upsilon$. In other words, the tightness of RUR is inversely proportional to the mixedness.

As we can see from Fig.2 the tightness and the mixedness of the steady state has nothing to do with the initial state we choose. Let $t \to \infty$ and make use of Eqs.(4),(9),(10),(11)and(13) one can acquire $U_{t\to\infty} = 2\Upsilon_{t\to\infty}$, that is to say, the tightness of RUR has a strict linear relationship with the mixedness at the case that the system is in steady state. The expressions and evolution of $U_{t\to\infty}$ and $\Upsilon_{t\to\infty}$ are given as

$$U_{t\to\infty} = \lim_{t\to\infty}(\Delta \hat{A}^2 \Delta \hat{B}^2 - \frac{1}{4}|\langle[\hat{A},\hat{B}]\rangle|^2) = 1 - \frac{1}{(1+2\lambda^2)^2} \quad (15)$$

$$\Upsilon_{t\to\infty} = \lim_{t\to\infty} \Upsilon = \frac{1}{2}(1 - \frac{1}{(1+2\lambda^2)^2}) \quad (16)$$

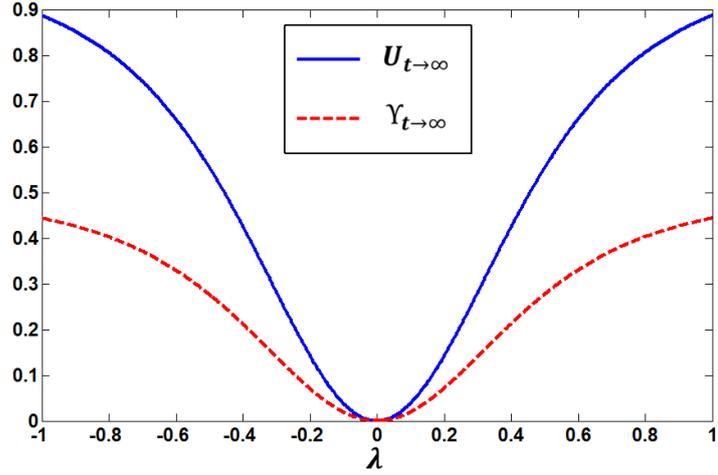

Fig.3. the evolution of $U_{t\to\infty}$ and $\Upsilon_{t\to\infty}$ with respect to $\lambda$, here $\beta = \pi/2$.

As shown in Fig.3, the tightness and mixedness are only affected by the value of $\lambda$, which represents the feedback strength. In addition, we can see that the tightness reaches the optimum value while the mixedness reaches the minimum value when the system has no feedback. Therefore, we can deduce that the feedback $\sigma_x$ can destroy the tightness of uncertainty in steady state. Contrary to the tightness, the mixedness can be enhanced by the feedback in steady state.

In order to study the changes of the tightness in the initial evolution of the system, we take $t = 0.5$. The evolution of the tightness and the mixedness with respect to feedback strength is shown in Fig.4.

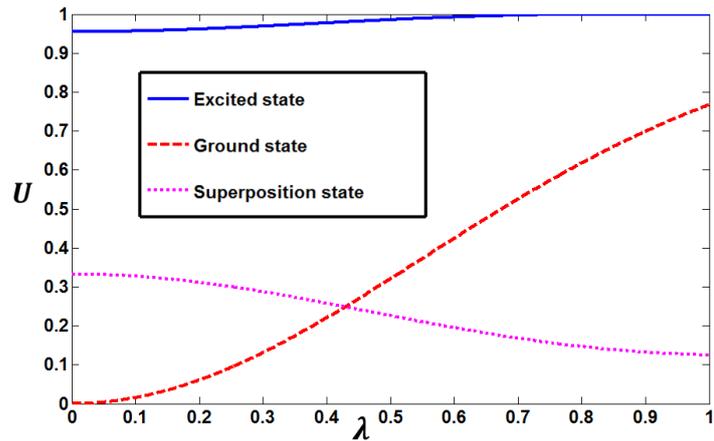

Fig.4. the evolution of $U$ with respect to $\lambda$, here excited state stands for $|e\rangle$, ground state stands for $|g\rangle$ and superposition state stands for $(|g\rangle + |e\rangle)/\sqrt{2}$, here $\beta = \pi/2$.

It can be seen from Fig.4 that the tightness can be enhanced by tuning the feedback $\lambda\sigma_x$ when we choose the superposition state rather than the ground and excited one as initial state at the beginning of the evolution.

The feedback $\lambda\sigma_y$ will be investigated in the following. Submitting $\beta = 0$ into Eqs. (9), (10), (11) and (13), one can obtain the evolution of $U$ and $\Upsilon$ with respect to time for different initial state in Fig.5, here we take $\lambda = 1$.

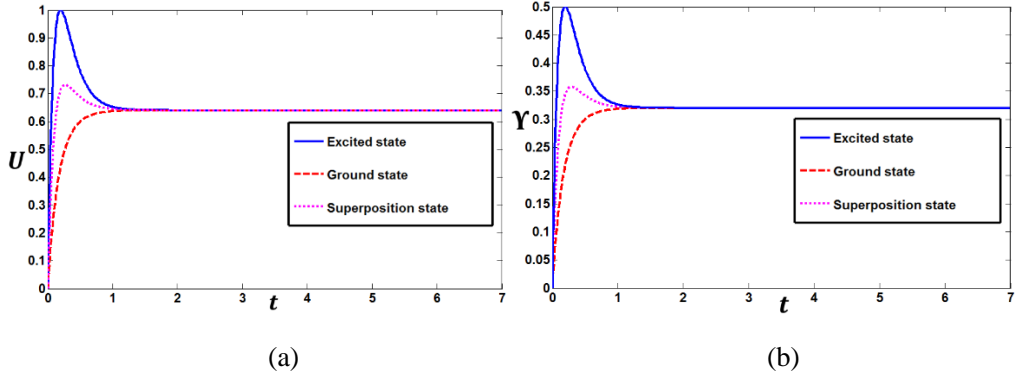

(a)    (b)

Fig.5. the evolution of $U$ in (a) and $\Upsilon$ in (b) with respect to $t$ for different initial state, here excited state stands for $|e\rangle$, ground state stands for $|g\rangle$ and superposition state stands for $(|g\rangle + |e\rangle)/\sqrt{2}$, here $\beta = 0$.

As shown in Fig.5, different form the feedback $\sigma_x$, when $\sigma_y$ is chosen as the feedback taking the ground state as initial one can be more effective to enhance the tightness of uncertainty. Meanwhile, exactly opposite to the tightness, the mixedness can be effectively improved by choosing excited state as initial state. In addition, the similar structure between (a) and (b) indicates the tightness of the RUR is inversely proportional to the mixedness of the system. At the same time, it is easy to find that the tightness and the mixedness has nothing to do with the initial state we choose when the system reach the steady state which is similar to the conclusion we obtain at the case that taking $\sigma_x$ as the feedback. Let $t \to \infty$, one can acquire $U_{t\to\infty} = 2\Upsilon_{t\to\infty}$. In other words, the tightness has a strict linear relationship with the mixedness in steady state at the case choosing $\lambda\sigma_y$ as feedback. The expressions and evolution of $U_{t\to\infty}$ and $\Upsilon_{t\to\infty}$ are given as

$$U_{t\to\infty} = \lim_{t\to\infty}(\Delta \hat{A}^2 \Delta \hat{B}^2 - \frac{1}{4}|\langle[\hat{A},\hat{B}]\rangle|^2) = \frac{4\lambda^2(1+\lambda)^2}{(1+2\lambda+2\lambda^2)^2} \quad (17)$$

$$\Upsilon_{t\to\infty} = \lim_{t\to\infty} \Upsilon = \frac{2\lambda^2(1+\lambda)^2}{(1+2\lambda+2\lambda^2)^2} \quad (18)$$

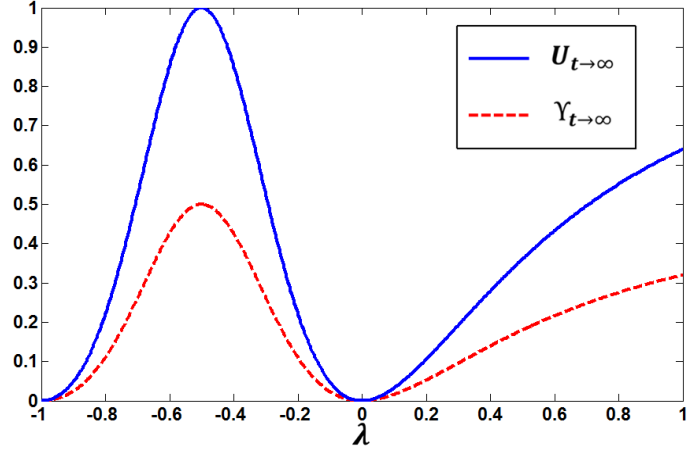

Fig.6. the evolution of $U_{t\to\infty}$ and $\Upsilon_{t\to\infty}$ with respect to $\lambda$, here $\beta = 0$

We can deduce from Eqs. (17) and (18) that the tightness and mixedness are only affected by the feedback strength. In addition, it can be seen from Fig.6 that the tightness of the uncertainty reaches the optimum value while the mixedness reaches the minimum value at the points $\lambda = 0$ and $\lambda = -1$. Therefore, we can deduce that the feedback can enhance the mixedness of the system but cannot improve the tightness of uncertainty in steady state.

Similar to Fig.4, submitting $t = 0.5$ into Eqs. (9), (10), (11) and (13), one can obtain the evolution of the tightness with respect to $\lambda$ for different initial state in Fig.7.

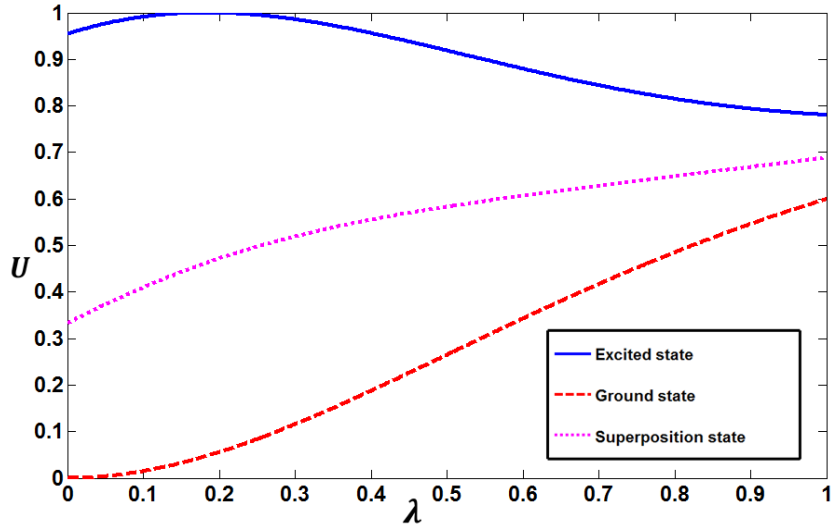

Fig.7. the evolution of $U$ with respect to $\lambda$, here excited state stands for $|e\rangle$, ground state stands for $|g\rangle$ and superposition state stands for $(|g\rangle + |e\rangle)/\sqrt{2}$, here $\beta = 0$.

It can be deduced from Fig.7 that the tightness can be enhanced by the $\lambda\sigma_y$ type feedback at the beginning of the evolution when we use the excited state as initial one.

We have investigated the feedback $\lambda\sigma_x$ and $\lambda\sigma_y$ respectively in above. The effect of the superposition of the two kinds of feedback $\lambda(\sin(\beta)\sigma_x + \cos(\beta)\sigma_y)$ on the tightness and mixedness will be studied in the following. Taking $t \to \infty$ and making use of Eqs. (9), (10), (11) and (13), one can obtain

$$U_{t\to\infty} = 1 - \frac{(1+2\lambda\cos(\beta))^2}{(1+2\lambda^2+2\lambda\cos(\beta))^2} \tag{19}$$

$$\Upsilon_{t\to\infty} = \frac{1}{2}(1 - \frac{(1+2\lambda\cos(\beta))^2}{(1+2\lambda^2+2\lambda\cos(\beta))^2}) \tag{20}$$

As shown in Eqs. (19) and (20), the conclusion that the tightness has a strict linear relationship with the mixedness in steady state still can be deduced for choosing the superposition feedback $\lambda(\sin(\beta)\sigma_x + \cos(\beta)\sigma_y)$. The evolution of the tightness is shown in Fig.8

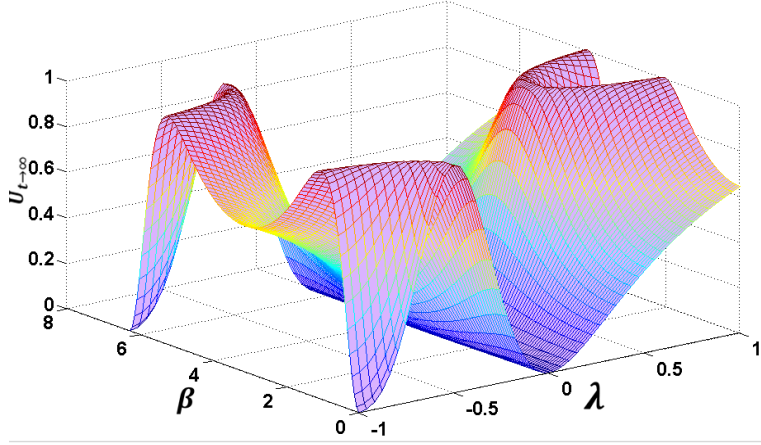

Fig.8. the evolution of $U_{t\to\infty}$ with respect to $\lambda$ and $\beta$

It can be seen from Fig.8 that there are several points where uncertain inequality becomes uncertain equality which predicates the tightness is optimum. It obviously that the points representing nonexistent of feedback belong to the points mentioned above. Therefore, we can deduce that the feedback still cannot enhance the tightness of uncertainty in steady state by choosing the superposition feedback.

### B. The feedback $\mu\sigma_z$

As a comparison, the feedback $\mu\sigma_z$ will be investigated in the following. Substituting $F = \mu\sigma_z$ into Eq. (3), one can obtain the function of density matrix:

$$\rho(t) = \begin{pmatrix} \rho_{11}(t) & \rho_{12}(t) \\ \rho_{12}^*(t) & 1 - \rho_{11}(t) \end{pmatrix} \tag{21}$$

with the elements

$$\rho_{11}(t) = e^{-t}\sin(\alpha)^2 \tag{22}$$

$$\rho_{12}(t) = \frac{e^{-t/2}\sin(2\alpha)}{2\Gamma} - i\frac{8e^{-t}(e^{t/2}-\Gamma)\mu\sin^2(\alpha)}{2\Gamma(1-4\mu^2)} \qquad (23)$$

where $\Gamma = \exp(2t\mu^2)$. After a simple calculation, we have

$$\left(\Delta\hat{A}\right)^2 = \frac{1}{2}e^{-t(1+4\mu^2)}(-1 + 2e^{t+4t\mu^2} + \cos(4\alpha)) \qquad (24)$$

$$\left(\Delta\hat{B}\right)^2 = 2e^{-2t}\sin(\alpha)^2(2e^t + \cos(2\alpha) - 1) \qquad (25)$$

$$\left|\langle[\hat{A},\hat{B}]\rangle\right|^2 = \frac{256e^{-2t}(-1+e^{\frac{t}{2}-2t\mu^2})^2\mu^2\sin(\alpha)^4}{\mathcal{R}} \qquad (26)$$

$$\left|\langle\{\check{A},\check{B}\}\rangle\right|^2 = 4e^{-t(3+4\mu^2)}[e^t - 2\sin(\alpha)^2]^2\sin(2\alpha)^2 \qquad (27)$$

$$\Upsilon = \frac{\sin(\alpha)^2\left(\mathcal{R}\Gamma(2e^t+\cos(2\alpha)-1)-2\left(e^t\mathcal{R}\cos(\alpha)^2+16(e^{t/2}-\Gamma)^2\mu^2\sin(\alpha)^2\right)\right)}{\mathcal{R}\Gamma e^{2t}} \qquad (28)$$

with $\mathcal{R} = (1-4\mu^2)^2$. Similarly, making use of Eqs. (24), (25), (26), (27) and (28), one can obtain $W = \Delta\hat{A}^2\Delta\hat{B}^2 - \left|\langle[\hat{A},\hat{B}]\rangle/2\right|^2 - \left|\langle\{\check{A},\check{B}\}\rangle/2\right|^2 = 2\Upsilon$, which means that there also exists a strict liner relationship between the tightness of SUR and the mixedness of the system when we choose the feedback $\mu\sigma_z$. In addition, for the RUR, taking advantage of the above formulas, we can acquire the evolution of $U$ and $\Upsilon$ with respect to time for different initial state in Fig.9, here we take $\mu = 1$.

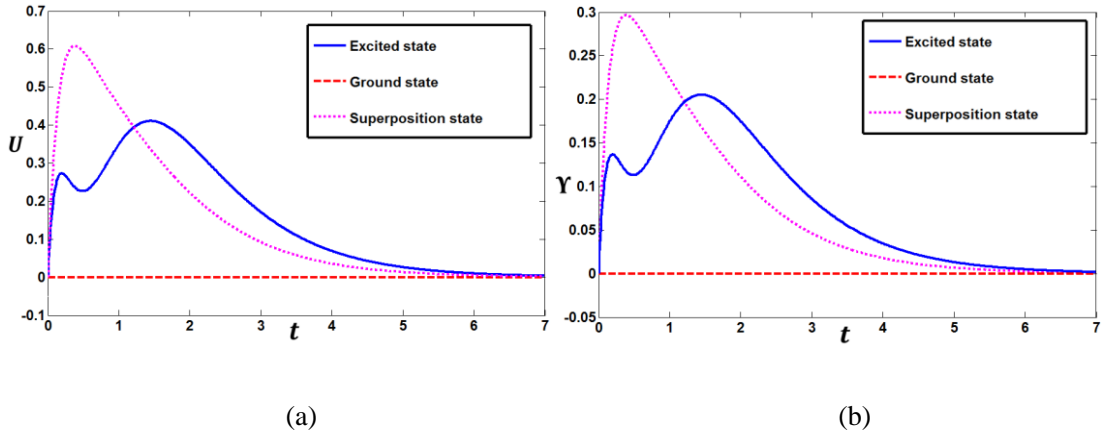

(a)          (b)

Fig.9. the evolution of $U$ in (a) and $\Upsilon$ in (b) with respect to $t$ for different initial state. Here excited state stands for $|e\rangle$, ground state stands for $|g\rangle$ and superposition state stands for

$$(|g\rangle + |e\rangle)/\sqrt{2}$$

As shown in Fig.9, the tightness and the mixedness of the steady state has nothing to do with the initial state we choose. Let $t \to \infty$, we have $U_{t\to\infty} = \Upsilon_{t\to\infty} = 0$, which means the tightness of the steady state has nothing to do with the feedback we choose. In addition we can deduce that $\lim_{t\to\infty} U/\Upsilon = 2$ which is the same as we get for the feedback $\lambda(\sin(\beta)\sigma_x + \cos(\beta)\sigma_y)$.

Similarly, taking $t = 0.5$, one can obtain the evolution of the tightness with respect to the strength of feedback in Fig.10

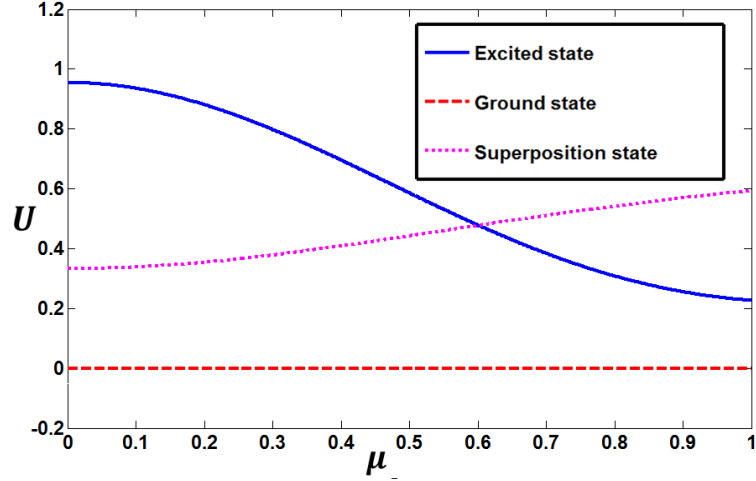

Fig.10. the evolution of $U$ with respect to $\mu$ and $\alpha$, here excited state stands for $|e\rangle$, ground state stands for $|g\rangle$ and superposition state stands for $(|g\rangle + |e\rangle)/\sqrt{2}$

We can obtain that tuning the feedback $\mu\sigma_z$ can improve the tightness at the beginning of the evolution only if we choose the excited state as the initial one. By comparing Fig.3, Fig.7 and Fig.10, we can find that the state with a better tightness can be obtained by tuning feedback when we choose the excited and superposition state as initial one at the beginning of the evolution.

## IV. CONCLUSIONS

In conclusion, we have investigated the effects of the different feedback types on the tightness of the uncertainties and the mixedness of the state in qubit system. For the SUR, we obtain that the tightness of SUR has a strict linear relationship with the mixedness $\Upsilon$ of the system which means the less the mixedness is the better the tightness of the SUR is, and the conclusion has nothing to do with the feedback we choose. As for the RUR, first of all, we can obtain the state with a better tightness at the beginning of the evolution by tuning the feedback. At the same time we find that the tightness of RUR can be destroyed by the feedback $\lambda(\sin(\beta)\sigma_x + \cos(\beta)\sigma_y)$ in steady state and cannot be affected by the feedback $\mu\sigma_z$. In addition, the evolution of the value of $U$ share a similar structure with the mixedness, that is to say there exists an inverse relationship between the tightness of RUR and the mixedness, what's more the relationship turns into a strict linear one when the system reach the steady state.

**Acknowledgments**

This work is supported by the National Natural Science Foundation of China (Grant No. 11574022).